# Macroscopic Invisibility Cloaking of Visible Light

Xianzhong Chen[1], Yu Luo[2], Jingjing Zhang[3], Kyle Jiang[4], John B. Pendry[2] and Shuang Zhang[1*]

[1]*School of Physics and Astronomy, University of Birmingham, Birmingham, B15 2TT, UK*

[2]*Department of Physics, Blackett Laboratory, Imperial College London, London, SW7 2AZ, UK*

[3]*DTU Fotonik - Department of Photonics Engineering, Technical University of Denmark, DK-2800 Kongens Lyngby, Denmark*

[4]*School of Mechanical Engineering, University of Birmingham, Birmingham, B15 2TT, UK*

*To whom the correspondence should be addressed: s.zhang@bham.ac.uk*

## Abstract

Invisibility cloaks of light, which used to be confined to the imagination, have now been turned into a scientific reality, thanks to the enabling theoretical tools of transformation optics and conformal mapping. Inspired by those theoretical works, the experimental realisation of electromagnetic invisibility cloaks has been reported at various electromagnetic frequencies. All the invisibility cloaks demonstrated thus far, however, have relied on nano- or micro-fabricated artificial composite materials with spatially varying electromagnetic properties, which limit the size of the cloaked region to a few wavelengths. Here we report realisation of a *macroscopic volumetric* invisibility cloak constructed from *natural* birefringent crystals. The cloak operates at *visible* frequencies and is capable of hiding three-dimensional objects of the scale of centimetres and millimetres. Our work opens avenues for future applications with macroscopic cloaking devices.

The pioneering theoretical works of Pendry and Leonhardt [1,2], fulfilling "invisibility cloaking", have inspired researchers around the world and have motivated many significant achievements[3-16], including the demonstration of invisibility cloaks at microwave[3-7], terahertz[8] and optical frequencies[9-12]. The invisibility-cloak design is enabled by transformation optics, using a coordinate transformation of the governing Maxwell's equations [17,18]. Such invariant transformations map a particular region in free space to a spatial domain, usually with position-dependent and anisotropic optical parameters. Notably, in 2003, Greenleaf et al. showed that a region in space could be rendered undetectable by electrical impedance tomography through invariant transformation of the Poisson's equation that describes the electrostatic potential.[19]

While the initial proposal for an invisibility cloak involves extreme values of material properties that normally require resonant structures leading to narrow-band operation[1,3], a special type of cloaking scheme, the so-called "carpet cloak", circumvents this issue by transforming a bulging reflecting surface into a flat one, hiding anything that is underneath the bulging surface from outside observers[15]. This carpet cloaking scheme has enabled the experimental realisation of cloaks at microwave, terahertz and even optical frequencies [4,7-10,12]. These demonstrations prove that broadband invisibility can be accomplished with realistic material properties. However, most of the invisibility cloaks demonstrated so far rely on spatially varying index profiles, which, in the optical regime, usually require nanofabrication processes (such as focus ion beam, electron-beam lithography and direct laser writing) to accurately define the patterns on the nanometre scale [9,10,12]. These time-consuming nanofabrication processes limit the overall cloaking region to a few wavelengths.

In this Letter, we report the demonstration of a macroscopic volumetric cloaking device operating at visible frequencies, which can conceal objects of sizes of at least three orders of magnitude larger than the wavelength of light in all three dimensions. The cloaking design uses birefringence in a natural crystal calcite, thus eliminating the necessity of time-consuming nanofabrication processes and enabling the realisation of cloaking at macroscopic scales. The cloaking effect was directly observed for red and green laser beams and incoherent white light without resorting to use of a microscope. The demonstration of a macroscopic invisibility cloak paves the way for future applications of invisibility cloaking.

The cloak is based on a recent theoretical work[20] that indicates that a carpet cloak can be achieved with spatially homogeneous anisotropic dielectric materials. A schematic of the triangular cloaking design is shown in Fig. 1, where a virtual space with a triangular cross section of height $H_2$ and filled with an isotropic material of permittivity $\varepsilon$ and $\mu$ ($\mu=1$) (blue region in Fig. 1a) is mapped to a quadrilateral region in the physical space with anisotropic electromagnetic properties $\varepsilon'$ and $\mu'$ (brown region in Fig. 1b). Thus, the cloaked region is defined by the small gray triangle of height $H_1$ and half-width $d$. Mathematically, the transformation is defined by

$$x' = x, \quad y' = \frac{H_2 - H_1}{H_2} y + \frac{d - x\,\mathrm{sgn}(x)}{d} H_1, \quad z' = z \qquad (1)$$

where ($x'$, $y'$, $z'$) and ($x$, $y$, $z$) correspond to the coordinates of the physical space and virtual space, respectively. Applying this coordinate transformation to Maxwell's equations, we obtain the corresponding electromagnetic parameters of the quadrilateral cloaking region:

$$\varepsilon' = \varepsilon \hat{M}, \quad \mu' = \hat{M} \qquad (2)$$

where

$$\hat{M} = \begin{pmatrix} \dfrac{H_2}{H_2 - H_1} & -\dfrac{H_1 H_2}{(H_2 - H_1)d}\mathrm{sgn}(x) & 0 \\ -\dfrac{H_1 H_2}{(H_2 - H_1)d}\mathrm{sgn}(x) & \dfrac{H_2 - H_1}{H_2} + \dfrac{H_2}{H_2 - H_1}\left(\dfrac{H_1}{d}\right)^2 & 0 \\ 0 & 0 & \dfrac{H_2}{H_2 - H_1} \end{pmatrix} \qquad (3)$$

It is worth noticing from Eqns. 2 and 3 that materials with spatially invariant optical properties can fulfil the requirement for this triangular cloak, in contrast to all other cloaking demonstrated to date. An ideal cloak requires both the electric permittivity and magnetic permeability to be anisotropic. On the other hand, it has been shown that a reduced cloaking scheme, namely one that works for a specific polarisation, can eliminate the necessity to engineer both the permittivity and permeability[3, 13]. Specifically, considering the case where light with transverse-magnetic-field (TM) polarisation is incident in the *x-y* plane, the relevant optical parameters would be $\varepsilon_{x-y}$ (tensor of permittivity in the *x-y* plane) and $\mu_z$, and as far as the trajectory of the light is concerned, the electromagnetic parameters can be simplified as

$$\varepsilon'_{x-y} = \varepsilon \begin{pmatrix} \left(\dfrac{H_2}{H_2 - H_1}\right)^2 & -\dfrac{H_1 H_2^2}{(H_2 - H_1)^2 d}\mathrm{sgn}(x) \\ -\dfrac{H_1 H_2^2}{(H_2 - H_1)^2 d}\mathrm{sgn}(x) & 1 + \left(\dfrac{H_2}{H_2 - H_1}\right)^2\left(\dfrac{H_1}{d}\right)^2 \end{pmatrix} \qquad (4)$$

$$\mu'_z = 1 \tag{5}$$

Thus, the reduced cloaking design described by Eqns. 4 and 5 can be realised using natural anisotropic materials (such as calcite and calomel), which are readily available in macroscopic sizes.

Here, the macroscopic invisibility cloak is designed using two calcite prisms glued together with the protruding bottom surface of the cloak serving as a deformed mirror, as shown in Fig. 1c. Calcite is a uniaxial birefringent crystal with refractive indices of ordinary ($n_o$) and extraordinary light ($n_e$) of approximately 1.66 and 1.49, respectively, at a wavelength of 590 nm. To meet the transformation requirement described by Eqn. 4, the optical axis of calcite prisms forms an angle γ with the *y*-axis, and the relation between the geometrical parameters of the cloak, the optical axis orientation γ, and the permittivity of the virtual space, ε, is given as follows:

$$\frac{H_1}{d} = -\frac{(n_e^2 - n_0^2)\sin\gamma\cos\gamma}{n_0^2\cos^2\gamma + n_e^2\sin^2\gamma} \tag{6}$$

$$\frac{H_1}{H_2} = 1 - \sqrt{\frac{n_0^2\sin^2\gamma + n_e^2\cos^2\gamma}{n_0^2\cos^2\gamma + n_e^2\sin^2\gamma} - \left(\frac{H_1}{d}\right)^2} \tag{7}$$

$$\varepsilon = \left(\frac{H_2 - H_1}{H_2}\right)^2 (n_0^2\cos^2\gamma + n_e^2\sin^2\gamma) \tag{8}$$

In the specific design shown in Fig. 1C, γ = 30°, the geometrical parameters are calculated using Eqns. 6 and 7 and are indicated in Fig. 1c. The cloaked region has a triangular cross section formed by the two bottom facets, each forming an angle of about 5° with the *x-z* plane. The minimum dimension of the cloaked region—the height of the triangle—is close to 1.2 mm, which is more than three orders of magnitude larger than the wavelengths at visible frequencies. The refractive index of the virtual space is 1.532 by Eqn. 8, indicating that the cloak will cause no deflection or translation of light if it is placed on a reflective ground plane and immersed in an optical medium with the same refractive index.

A ray tracing calculation was performed to verify the design described above. As shown in Fig. 2a and 2b, beams with TM polarisation incident at two different angles are both reflected without distortion, numerically verifying the cloaking design given by Eqns. 6-8. On the other hand, for TE polarisation (which does not exhibit a cloaking effect), the light is reflected into different directions by the two bottom facets, leading to significant splitting, which can be observed in the far field (Fig. 2c and 2d).

To demonstrate the performance of the cloak in the visible range, we first characterised the calcite cloak in air using a green laser at wavelength of 532 nm (Fig. 3a, left). Even without a metal film covering the bottom surface of the cloak, light experiences total internal reflection for a large range of incident angles. To better visualise the effect of the cloaking, a mask with an arrow pattern was placed in front of the laser head such that the emitted laser beam contains the same pattern (Fig. 3a, right), which can be used to characterise the distortion of the beam after being reflected by the triangular bump covered by the cloak. In addition, the laser beam goes through a linear polariser that controls the polarisation of the beam to be either transverse electric (TE) or transverse magnetic (TM); here, the TE polarisation serves as the control sample that shows the results without the cloaking effect. The reflected beam is projected on a screen approximately 18 centimetres away from the prism, and the projected image is captured by a camera. As a reference, the image of the laser beam reflected by a flat mirror was shown in Fig. 3b , which contained a horizontally flipped arrow pattern.

Because the bottom surface of the cloak consists of two flat facets forming an angle about 10°, the reflection from the bottom bulging surface would split the laser beam into two for the TE polarised beam, which does not experience the cloaking effect, forming an angle of 20° inside the prism and greater than 30° outside the prism. This effect is clearly observed on the image projected on the screen for the green laser beam, as shown in Fig. 3c. The angle formed between the two split beams is consistent with a simple ray tracing analysis.

On the other hand, with TM polarisation, the reflected beam projected on the screen shows no splitting at all (Fig. 3d), in stark contrast to the case with TE polarisation. Although the cloak is designed for a wavelength of 590 nm, due to the latitude in manufacturing the prisms, the geometric parameters slightly deviate from the design. This effect leads to an almost perfect cloaking at $\lambda$=532 nm (green light) instead. A small dark stripe appears in the centre of the pattern due to the imperfection in the alignment of the two calcite crystals, resulting in scattering at the apex of the triangular cloaked region. An invisibility cloak should hide objects at all incident angles. This cloaking effect is confirmed by the measurements at three different incident angles (39.5°, 64.5°, 88°) with mixed TE and TM polarisations, as shown in Fig. 3 (e, f, g), in which the central image corresponds to the TM polarisation and the two projected arrow segments images far from the centre correspond to the TE polarisation. The projected images of the TM polarised beam at all incident angles show no distortion of the pattern, serving as direct evidence that the calcite cloak transforms the protruding bottom surface into a flat mirror.

We next demonstrate the performance of the cloak for red light using a laser that emits light at wavelengths ranging from 630 nm to 680 nm with a spectral peak at λ=650 nm. As shown in Fig. 3h, there is a small gap in the central image (TM polarisation). This gap can be attributed to the optical dispersion of calcite crystal, such that the cloaking condition is not rigorously fulfilled at these wavelengths. Consequently, the pattern is slightly different from that reflected from a flat surface (Fig. 3i). Nonetheless, in comparison with the large separation between the two segmented images of the uncloaked TE polarisation, the gap size (2 mm) of the TM polarisation is about two orders of magnitude less. To an outside observer, this difference corresponds to a virtual reflective surface consisting of two facets, each forming an angle of only around 0.1° with the horizontal (x-z) plane. This result, together with that of green light, indicates that the cloak covers at least a 100 nm bandwidth at visible frequencies.

The broadband operation of the macroscopic calcite cloak is further supported by the imaging of white-coloured alphabetic letters (from A to Z, flipped horizontally) printed on a sheet of black paper reflected by the cloak system. A schematic of the measurement is shown in Fig. 4a. Note that the bottom surface is now coated with Ti (2 nm)/Ag (200 nm)/Au (50 nm) to reflect light for all incident angles. With TE polarisation (Fig. 4b), due to the large splitting caused by the mirror deformation, the image collected by the camera consists of letters from two largely separated locations ('C', 'D', 'R' to 'U'). On the other hand, switching the polarisation to TM leads to imaging of five consecutive letters from the same location (from 'H' to 'L'), as if the bottom surface of the cloak were flat (Fig. 4c). The overall cloaking effect is striking, notwithstanding the presence of a rainbow at the edge of the letters arising from the dispersion of calcite crystal. Because white light covers the whole visible spectrum, this finding unambiguously demonstrates the broadband operation of our calcite cloak.

The dispersion of calcite has a slight effect on the cloaking performance, which was shown by the measurement with red laser beam in Fig. 3h and the rainbow edge effect in Fig. 4c. A ray tracing calculation was carried out to quantitatively analyse the influence of calcite dispersion on the cloaking performance. A beam with TM polarisation, once reflected by the cloak, is generally split into two except at the wavelength where the cloaking condition is rigorously satisfied. The two reflected beams deviate from that reflected by a horizontal flat mirror by an angle of $\phi_1$ and $\phi_2$, respectively, and the angle formed by the two beams is $\phi_1 - \phi_2$. Fig. 5a shows the ray tracing results for a cloak with the design parameters indicated in Fig. 1c and an incident angle θ=64.5°. The cloak works perfectly at the design wavelength close to 590 nm, which is confirmed by the ray tracing calculation. The deviation angles $\phi_1$, $\phi_2$ of the reflected beams are less than 1° across the visible range from 400 nm to 700 nm.

Due to the fabrication latitude, the cloaking condition was shifted to 532 nm, as shown by the measurements with green laser beam. Therefore, a direct comparison between the ray tracing calculation using the design parameters and the experimental observation would not be meaningful. Instead, we slightly modified the geometry of the cloak in the ray tracing to match the experimental observation of perfect cloaking at 532 nm, with the results given in Fig. 5b. At $\lambda = 650$ nm, the splitting angle formed by the two beams is $\phi_1 - \phi_2 = 0.68°$, which is consistent with our experimental observation for the red laser beam, namely, a separation of 2 mm on a screen 18 cm away from the prism, corresponding to a splitting angle of 0.64°.

It is interesting to note from Fig. 4 that the rainbow effect for TE polarisation is much less than that for TM polarisation. The ray tracing calculation verifies that the dispersion effect for TE is 2 orders of magnitude less than TM (Fig. 5c), which is consistent with our experimental observations.

An interesting question is whether any reflection exists at the interface between the two constituent calcite crystals with different orientations of the optical axis. It has been shown by Zhang et al. that for two contacting anisotropic crystals with their optical axis symmetric about the interface (as in our cloaking configuration), there is indeed no reflection at the interface regardless of the direction of light[21]. On the other hand, because we have adopted a reduced form of cloaking design (Eqns. 4, 5), a small impedance mismatch is introduced between the surrounding optical medium and the cloak, which results in a small amount of reflection at their interface. Nonetheless, this reflection can be eliminated if the interface is nano-structured in such a way that the permittivity undergoes an adiabatic transition from the calcite crystal and the surrounding medium, and the cloak itself would become completely invisible[22, 23, 24].

In summary, we have demonstrated the macroscopic cloak operating at visible frequencies, which transforms a deformed mirror into a flat one from all viewing angles. The cloak is capable of hiding three dimensional objects 3 - 4 orders of magnitudes larger than optical wavelengths, and therefore it satisfies a layman's definition of an invisibility cloak: namely, the cloaking effect can be directly observed without the help of microscopes. Because our work solves several major issues typically associated with cloaking: size, bandwidth, loss [25, 26] and image distortion[27], it paves the way for future practical cloaking devices.

## Acknowledgement:

We thank J. M. F. Gunn for insightful discussions and a critical reading of the manuscript. This work was supported by the University of Birmingham, HEFCE, EPS, and the EU project PHOME (Contract No. 213390). Y.L. acknowledges the financial support of the "Lee Family Scholarship" from Imperial College London.

Figure 1

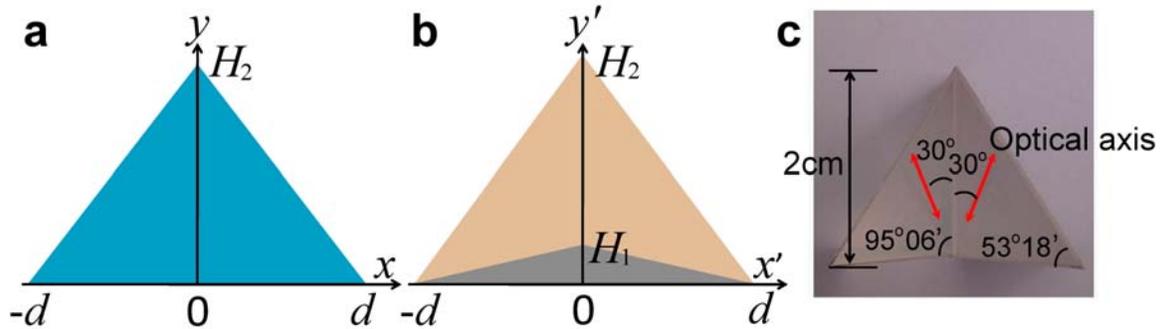

**Figure 1 Illustration of the transformation from real to virtual space and cloaking design.** In the transformation, a triangular cross section in a virtual space (**a**) filled with isotropic materials is mapped to a quadrilateral (brown region in (**b**)) with uniform and anisotropic optical properties. The cloaked region is defined by the small gray triangle wherein objects can be rendered invisible. **c.** A photograph of the triangular cloak, which consists of two calcite prisms glued together, with the geometrical parameters indicated in the figure. The dimension of the cloak along *z* direction is 2 centimetres. The optical axis, represented by red arrows, forms an angle of 30° with the glued interface.

Figure 2

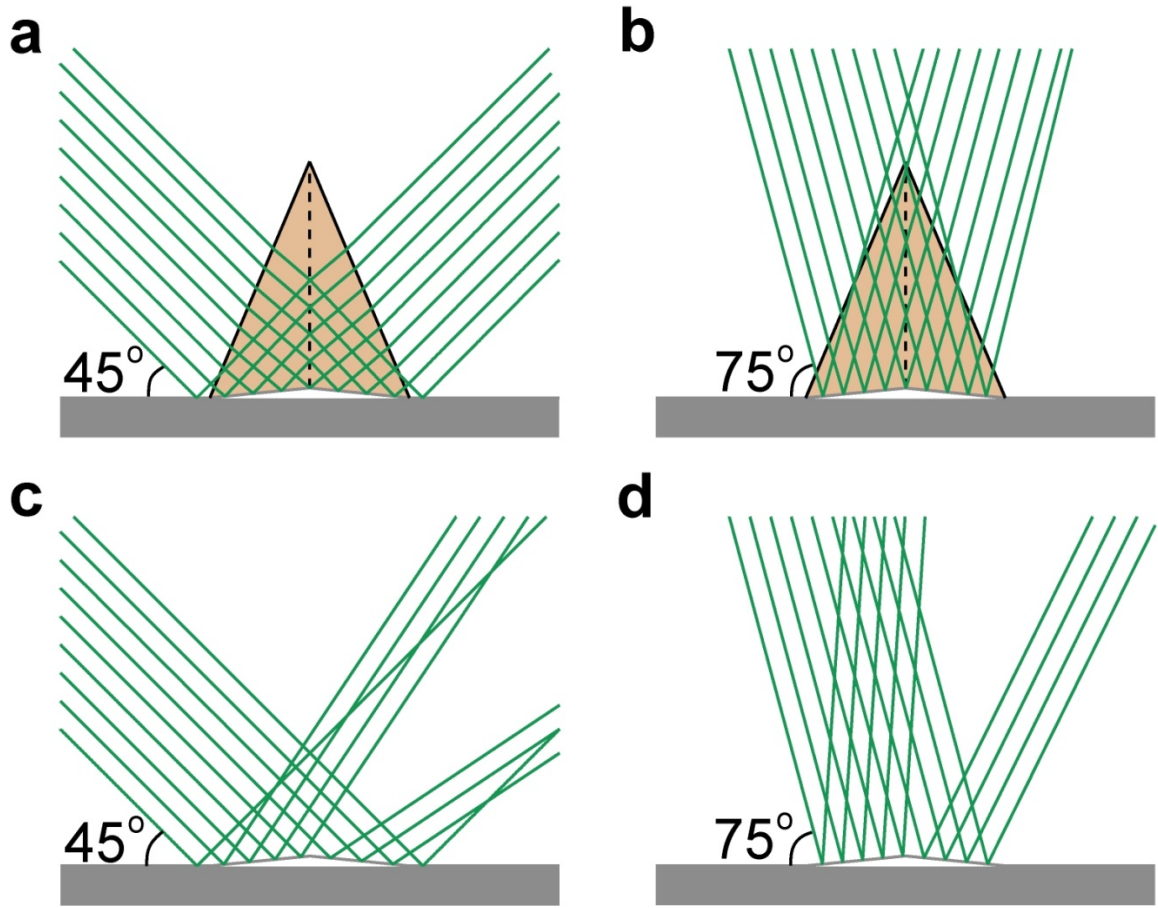

**Figure 2 Ray tracing of light passing through the cloak system at 532 nm.** The refractive index of the surrounding media is assumed to be 1.532. **a, b.** Light incident upon a cloaked bump at incident angles of (**a**) 45° and (**b**) 15°, showing almost no scattering. **c, d.** Without cloaking, the bump strongly scatters the light into different directions. Although the cloak is designed at 590 nm, however, in accordance with the experiment, the ray tracing calculation is performed at 532 nm.

Figure 3

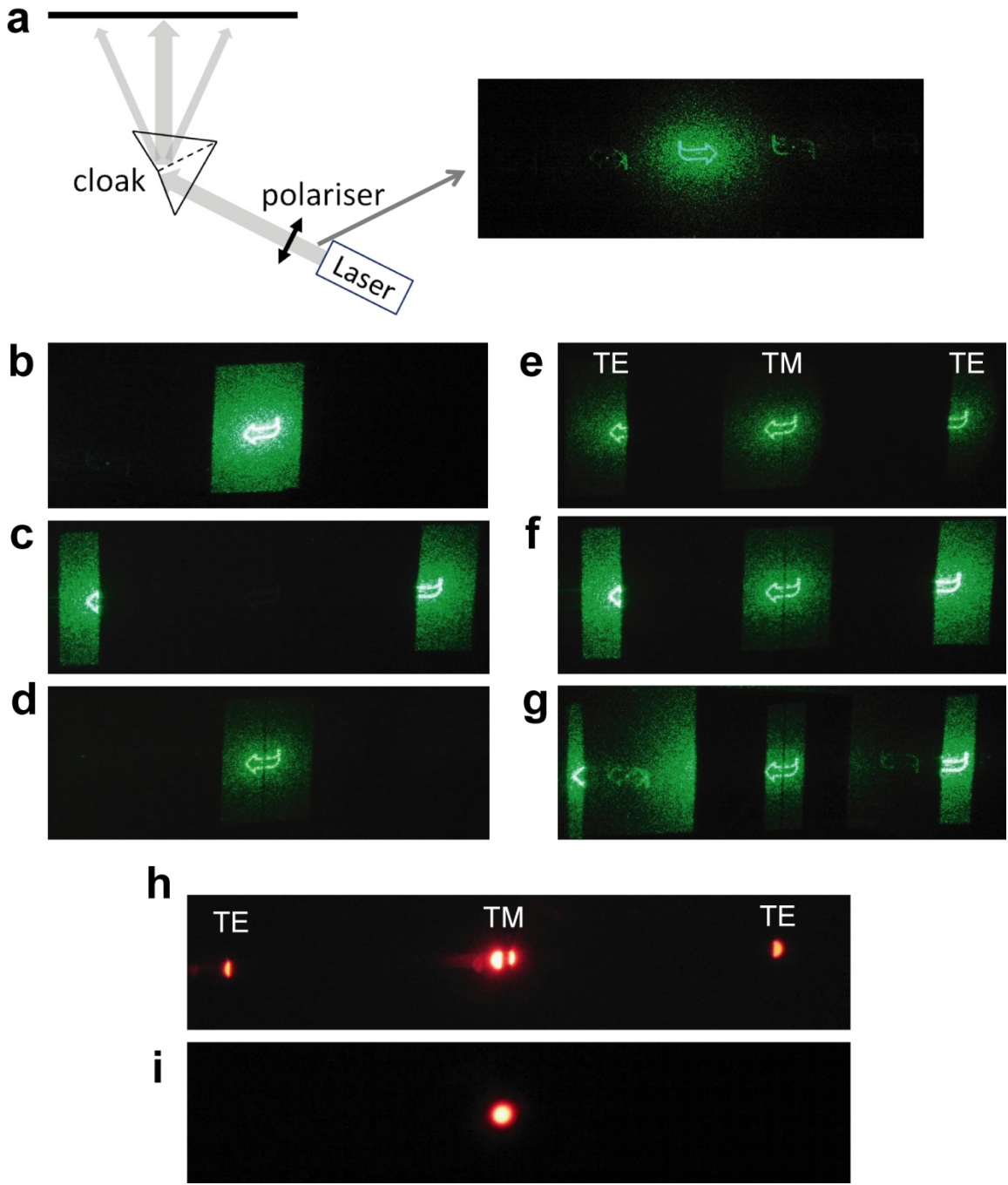

**Figure 3 Optical characterisation of the cloak using green and red laser beams. a.** (left) Schematic of the experimental setup. A patterned laser beam is reflected by a calcite cloak (or a flat reflective surface as control sample) and projected onto a screen. (right) The original pattern of the laser beam, which consists of a bright arrow in the centre and a number of flipped dim arrows on both sides. **b.** The pattern of the laser beam as reflected by a flat surface. The size of the mirror only allows the central arrow to be reflected. The projected arrow image is about 1.2 centimetres long in the horizontal direction. **c, d.** The projected image of the laser beam reflected by the calcite cloak for TE and TM polarisations, respectively. The TM measurement shows that the laser beam is not distorted by reflection by the triangular protruding surface. **e, f, g.** the projected images for mixed TE and TM polarisations at incidence angles of 39.5°, 64.5° and 88° respectively. For all incident angles, the central TM images are not distorted, the cloaked reflective bump appears to be a flat mirror to outside observers. Due to the limited size of the reflective surface, only the central arrow was reflected and subsequently changed its propagation direction, generally causing a large separation between its image projected on the screen and the others. However, in Fig. 3(**g**), for an incident angle close to the grazing angle, the change of direction is very small; therefore, the images of the reflected central arrow and the other dimmer arrows all appear in the field of view of the camera. **h, i.** The photographs of a red laser beam with mixed TE and TM polarisations projected on the screen after being reflected by (**h**) calcite cloak and (**i**) a flat surface, respectively, at an incident angle of 64.5°.

Fig. 4

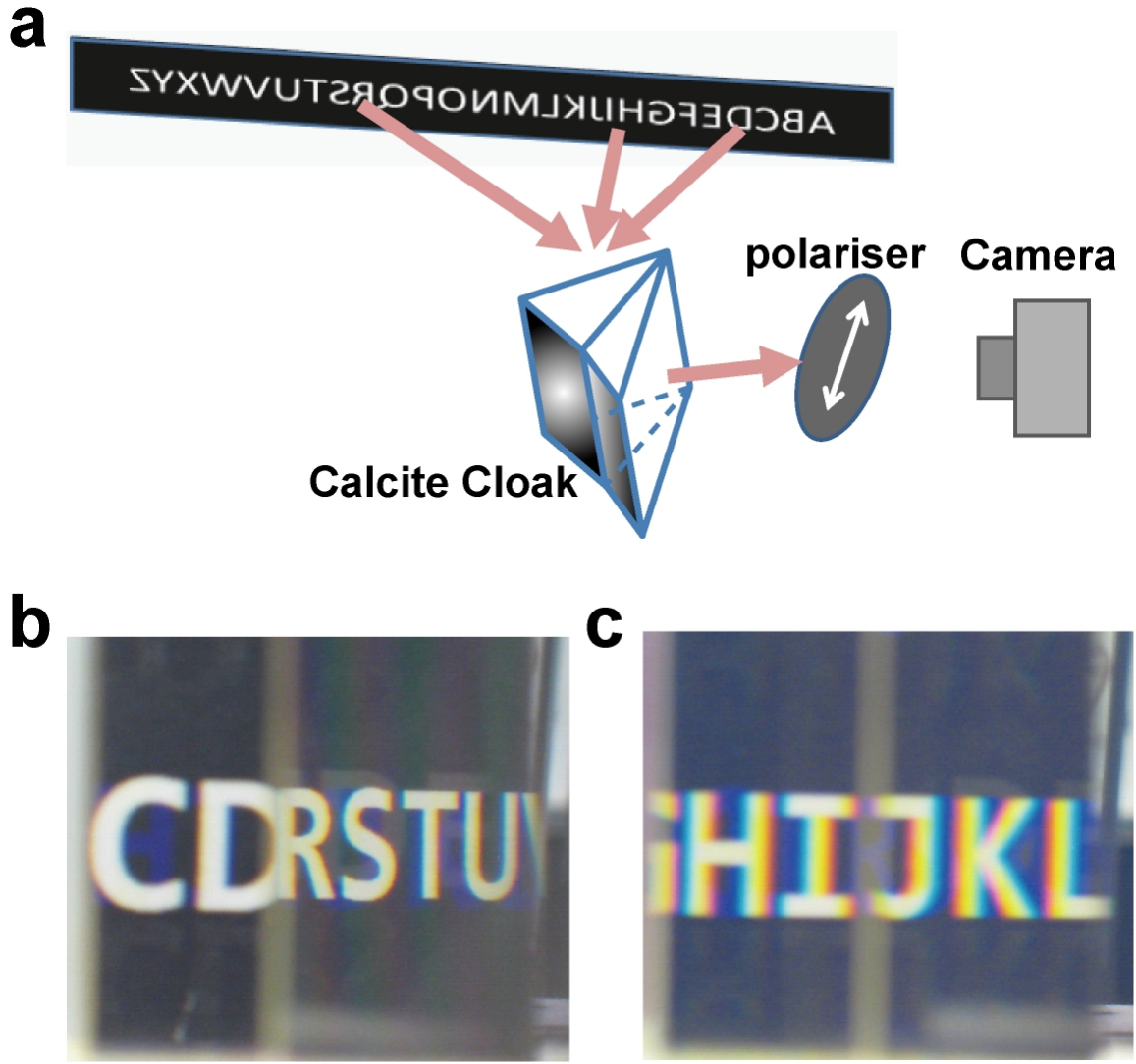

**Figure 4 Reflection of white alphabetic letters by the cloak. a.** Schematic of the optical setup. **b, c.** The reflected image captured by the camera for TE (without cloaking) and TM polarisations, respectively. Note that the rainbow appears at the edge of each letter for TM polarisation due to the optical dispersion of the calcite crystal.

Fig. 5

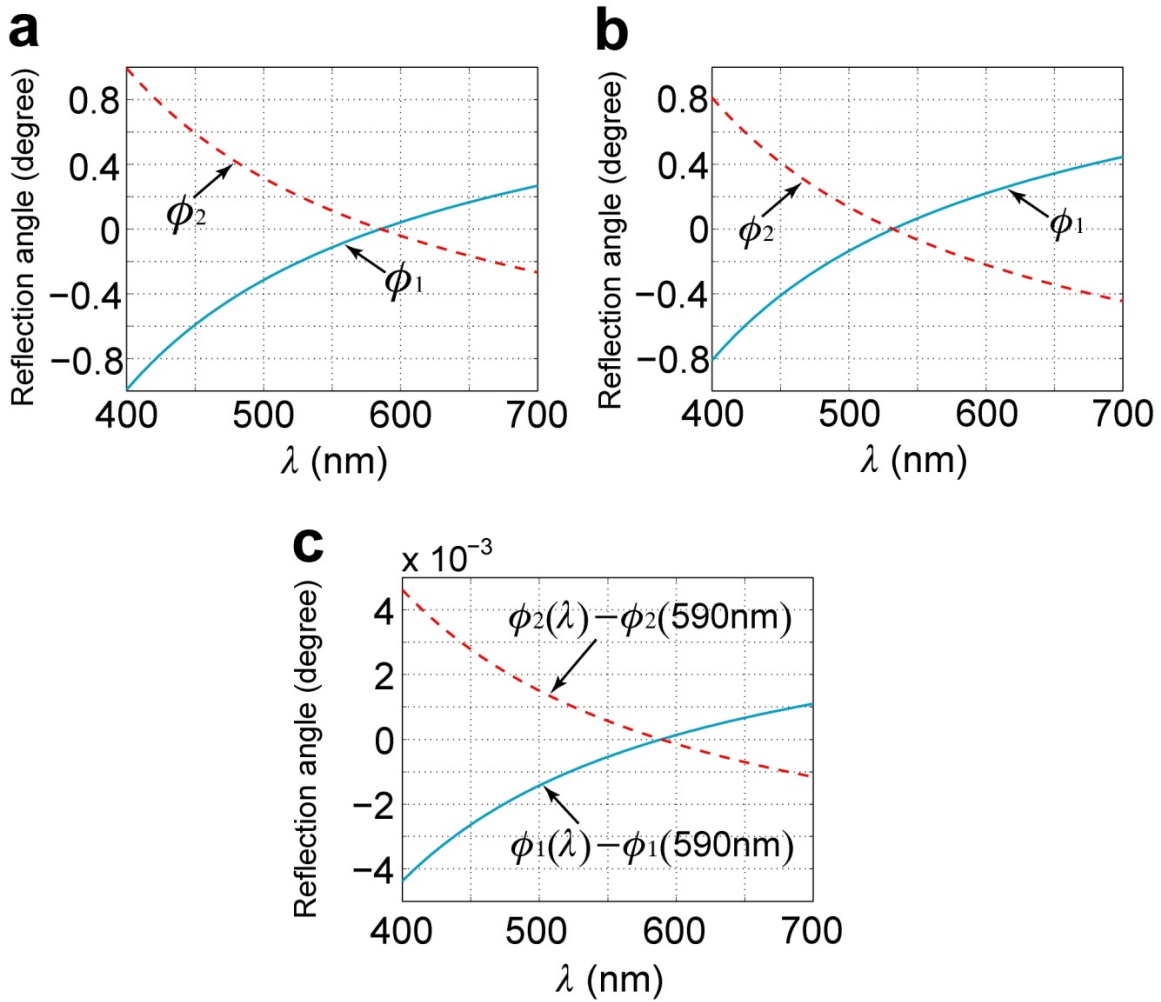

**Figure 5 Effect of Calcite dispersion on cloaking calculated by ray tracing a.** The deviation angles of the beams with TM polarisation reflected by the cloak with the design parameters indicated in Fig. 1c. **b.** same as (**a**), but for a slightly modified cloak geometry so that perfect cloaking occurs at wavelength of 532 nm, consistent with the experimental observations for the green laser beam. **c.** The effect of dispersion for TE polarisation in terms of the variation of reflection angles relative to those at $\lambda$ =590 nm. In all the figures, the incident angle was set to 64.5°.